\documentclass{ws-procs9x6-cpt16}
\begin{document}

\newcommand{\refeq}[1]{(\ref{#1})}
\def\etal {{\it et al.}}

\title{Is There a Signal for Lorentz Noninvariance \\ 
in Existing Radioactive Decay Data?}

\author{M.J.\ Mueterthies,$^1$  
D.E.\ Krause,$^{2,1}$ 
A.\ Longman,$^{1}$   
V.E.\ Barnes,$^{1}$ \\
and E.\ Fischbach$^{1}$ }

\address{$^1$Department of Physics and Astronomy, Purdue University\\
West Lafayette, IN 47907, USA}

\address{$^2$Department of Physics, Wabash College, 
Crawfordsville, IN 47933, USA}

\begin{abstract}
Measurements of the beta decay rates of nuclei have revealed annual periodicities with approximately the same relative amplitude even though the half lives range over nine orders of magnitude.  Here we show that this can be explained if the emitted neutrinos behave as if they propagate in a medium with a  refractive index which varies as the Earth orbits the Sun.  This refractive index may be due to fundamental Lorentz noninvariance, or apparent Lorentz noninvariance arising from interactions with solar or relic neutrinos, or dark matter.  Additionally, this medium could have consequences for experiments attempting to measure the neutrino mass.
\end{abstract}

\bodymatter

\phantom{}\vskip10pt\noindent
Recent studies of the decay rates $\lambda$ of radioactive nuclides have reported evidence for periodic variations superimposed on the familiar exponential decay laws.\cite{Jenkins 2009,SSR 2009,OKeefe 2013,Sturrock 2014,Sturrock 2015}  Although the most typical frequencies in $\lambda$ are $\sim 1$/yr and 11--13/yr, other periodicities have also been reported.  At this stage we cannot exclude the possibility that some of the annual signals may arise from seasonal variations in the response of various detectors to fluctuations in ``environmental'' factors such as temperature, humidity, and pressure.  However, similar considerations would  have much less impact on other observed periodicities, and hence will we will assume in this treatment that all the observed periodicities may be evidence of interesting new physics.  In what follows we suggest that a common feature seen in most of the periodic variations may be a hint to the presence of an effective Lorentz noninvariance (LNI).


\begin{table}
\tbl{Data from beta decay experiments (including $^{226}$Ra in equilibrium with its daughters) exhibiting annual periodicities. Here the mode $\beta^{-} (\beta^{+})$ indicates the emission of an electron (positron), and EC is electron capture. The half lives $\tau_{1/2}$ and decay $Q$ values are from S.Y.F.\ Chu \etal.\cite{IsotopeTable}}
{\begin{tabular}{@{} l r c c c c c  @{}}\toprule
Experiment & Source & Mode & Duration & $\tau_{1/2}$ (d) & $Q$ (keV) & $10^{3}\xi$ \\
\colrule
   Ellis\cite{Ellis,Parkhomov} & $^{56}$Mn & $\beta^{-}$ & 1978-87 & $1.1\!\times\!10^{-1}$ & 3695.5 & 3\\
      Purdue\cite{FischbachMn} & $^{54}$Mn & EC & 2008-13 & $3.1\!\times\!10^{2}$ & 1377.1 & $1$\\ 
   Parkhomov\cite{Parkhomov} & $^{60}$Co & $\beta^{-}$ & 1999-03 &$1.9\!\times\!10^{3}$& 2823.9 & 2 \\
  Norman\cite{OKeefe 2013,Norman} & $^{22}$Na/$^{44}$Ti & $\beta^{+}$,EC & 1994-96 & -- & -- & 0.34\\
   & $^{22}$Na & $\beta^{+}$ &   &$9.5\!\times\!10^{2}$& 2842.2 & -- \\
   & $^{44}$Ti & EC &   &$2.2\!\times\!10^{3}$& 267.5 & -- \\
  Schrader\cite{Schrader} & $^{154}$Eu & $\beta^{-}$ & 1990-96 & $3.1\!\times\!10^{3}$ & 1968.4 & 1\\
  Schrader\cite{Schrader} & $^{85}$Kr & $\beta^{-}$ & 1990-96 & $3.6\!\times\!10^{3}$ & 687.1 & 1\\
  Falkenberg\cite{Falkenberg} & $^3$H & $\beta^{-}$ & 1980-82 &$4.5\!\times\!10^{3}$& 18.59 & 3.7\\
  Schrader\cite{Schrader} & $^{152}$Eu & $\beta$, EC & 1990-96 &$4.9\!\times\!10^{3}$& 1874.3 & 1\\
    Parkhomov\cite{Parkhomov} & $^{90}$Sr & $\beta^{-}$ & 2000-10 &$1.1\!\times\! 10^{4}$& 546.0 & 1.3\\
     BNL\cite{BNL} & $^{32}$Si & $\beta^{-}$ & 1982-86 &$5.5\!\times\!10^{4}$& 224.5 & 1.5\\
  Schrader\cite{Schrader} & $^{108m}$Ag & $\beta^{+}$ & 1990-96 & $1.5\!\times\!10^{5}$ & 1918 & 1\\
     PTB\cite{PTB} & $^{226}$Ra & various & 1981-96 &$5.8\!\times\!10^{5}$& various & 1.5\\
  Mathews\cite{Mathews} & $^{14}$C & $\beta^{-}$ & 2016 & $2.2\!\times\!10^{6}$ & 156.4 & 2--4\\
  BNL\cite{BNL}  & $^{36}$Cl & $\beta^{-}$ & 1982-86 & $1.1\!\times\! 10^{8}$& 708.6 & 1.5\\
  Ohio Sate \cite{Ohio State} & $^{36}$Cl & $\beta^{-}$ & 2005--2011 &  $1.1\!\times\! 10^{8}$ & 708.6 & 5.8\\
\end{tabular}
}
\label{data table}
\end{table}

Table~\ref{data table} presents a summary of recent data ordered in terms of increasing half lives $\tau_{1/2}$, along with the corresponding $Q$ values for the decays, and the fractional change $\Delta\lambda/\lambda$ for the corresponding annual variation.  We see that even though values of $\tau_{1/2} = \ln 2/\lambda$ range over a factor of $\sim 10^{9}$, the fractional periodic variations of $\Delta\lambda/\lambda$ agree within roughly an order of magnitude.  More specifically, the data are well characterized by 
\begin{equation}
\frac{\Delta\lambda(t)}{\lambda} \simeq \xi\cos\left(\frac{2 \pi t}{T} - \phi\right),
\label{Delta lambda}
\end{equation}
where $T \simeq 1$~year and $\xi \sim 10^{-3}$.
This is surprising given that the data come from different experiments using a variety of detection systems,\cite{OKeefe 2013} which presumably have different levels of sensitivity to  environmental influences.  As but one example, we see from Table~\ref{data table} (and Fig.\ 3 of Ref.\ \refcite{Jenkins 2009}), that $\xi$ for $^{226}$Ra (in equilibrium with its daughters) obtained at PTB using a $4\pi$ $\gamma$-detector is very nearly the same as $\xi$ for the Si/Cl ratio obtained at BNL at the same time using a gas proportional counter.  This apparent consistency of $\xi$ is the basis for our proposed connection between radioactive decays and LNI.


We now show that the apparent universality of $\xi$ could be explained if the neutrinos involved act as if they propagate in a medium with a refractive index $n$ that depends on the Earth's motion around the Sun.   A number of circumstances in which neutrinos behave as if they are in a medium have been treated in recent years.  Such a description has been used to describe neutrino oscillations in matter, relic neutrinos, and LNI.  Here we will assume the simplest possibility, that the neutrinos experience a dispersion-free (i.e., energy-independent) refractive index $n = 1 + \epsilon$, where $0 < \epsilon \ll 1$, such that the wave number characterizing a neutrino matter wave is given by $k = nk_{\rm vac}$, where $k_{\rm vac}$ is the wave number in vacuum.  Since the angular frequency $\omega = kc/n = k_{\rm vac}c$,  the neutrino energy and momentum relations $E_{\nu} = \hbar\omega$ and $p_{\nu} = \hbar k$ then give
\begin{equation}
E_{\nu} \simeq pc/n \simeq (1-\epsilon)p_{\nu}c,
\label{E nu}
\end{equation}
where we assume that the neutrino mass $m_{\nu} \ll p_{\nu}/c$.  

One can show that the Lorentz violating dispersion relation given by Eq.~(\ref{E nu}) modifies the rates of beta decay and electron capture such that the fractional change $\Delta\lambda/\lambda \simeq 3\epsilon$, a constant, which is independent of the nucleus.  A simple way to see this is to note that the phase space for beta decays is proportional to $d^{3}k = n^{3}d^{3}k_{\rm vac}$, which implies that the decay rate in the medium is $\lambda = n^{3}\lambda_{\rm vac} \simeq (1 + 3\epsilon)\lambda_{\rm vac}$.  (Here we assume that the medium does not affect the matrix elements for the decay process.)

Let us now assume that the neutrino index of refraction depends on the position of the Earth.  If we write the Earth-Sun distance $r_{\oplus}(t)$ as
$r_{\oplus}(t) \simeq \overline{r}_{\oplus}\left[1 + \varepsilon_{\oplus}\cos\left(2\pi t/T_{\oplus} - \phi_{\oplus}\right)\right],
$ where $\overline{r}_{\oplus}$ is the average separation,  $\varepsilon_{\oplus} \simeq 0.0167$ is the orbital eccentricity, $T_{\oplus} =$~1 yr, and $\phi_{\oplus}$ is a phase, one can show that the beta decay rates will vary as
\begin{equation}
\frac{\Delta\lambda(t)}{\lambda(\overline{r}_{\oplus})} = \xi\cos\left(\frac{2\pi t}{T_{\oplus}} - \phi_{\oplus}\right),
\end{equation}
where $\xi$ is a constant which incorporates $\varepsilon_{\oplus}$ and the properties of the medium.  Thus, this model would describe an annual variation of beta decay rates with a constant amplitude for all nuclei, which is consistent with the observations which have $\xi \sim 10^{-3}$ as given in Table~\ref{data table}. 

 Of course, this is the simplest possible model, which doesn't include dispersion or address the nature and origin of the neutrino refractive index needed to explain these results.  If such an effective medium exists, it might be due to dark matter, relic neutrinos, solar neutrinos, or intrinsic LNI.  The model suggests the need for more observations to search for annual variations in beta decays over a wider range of nuclei since it predicts that the variations of the relative magnitude should be of order $10^{-3}$.


Finally, we note that the model described above could have significant consequences for experiments attempting to determine the neutrino mass $m_{\nu}$, such as KATRIN, which uses $^{3}$H, a nucleus exhibiting annual periodicity.\cite{Falkenberg}  As shown in Ref.\ \refcite{PDG}, most recent experimental determinations of $m_{\nu}^{2}$ lead to negative values.  These results can be accounted for using the refractive index model, which would modify the dispersion relation for massive neutrinos to be
\begin{equation}
E_{\nu}^{2} \simeq (1 - \epsilon)^{2}c^{2}p_{\nu}^{2} + m_{\nu}^{2}c^{4} \equiv c^{2} p_{\nu}^{2} + M_{\nu,{\rm eff}}^{2}c^{4}
\Rightarrow M_{\nu,{\rm eff}}^{2} \simeq m_{\nu}^{2} - 2\epsilon c^{2}p_{\nu}^{2}.
\end{equation}
Thus, the effective neutrino mass $M_{\nu,{\rm eff}}^{2} < 0$ when $2\epsilon c^{2}p_{\nu}^{2} > m_{\nu}^{2}$.  Combining this with the lower bound on the neutrino mass, $m_{\nu}\gtrsim 0.4$~eV/$c^{2}$, obtained from calculations of the stability of neutron stars in the presence of virtual neutrino-antineutrino exchange forces,\cite{Fischbach 1996} one finds that $M_{\nu,{\rm eff}}^{2}\simeq 0$ when $p_{\nu}c \simeq 10$~eV.  Thus, the neutrino refractive index model shows that $M_{\nu,{\rm eff}}^{2}$ measured can be positive, negative, or even zero, while the true neutrino mass squared $m_{\nu}^{2}$ is positive.  If this is true, it may be quite difficult to extract the true neutrino mass from measurements of the tritium decay spectrum.

\section*{Acknowledgments}
We thank Prof.\ Scott Mathews for making available to us his pre-publication results for $^{14}$C.

\end{document}